\documentclass[9pt,twocolumn,twoside]{pnas-new}

\usepackage{ulem}
\RequirePackage{color}

\newcommand{\red}[1]{{\color{red}{#1}}}

 \definecolor{MyDarkGreen}{rgb}{0.02,0.60,0.06}

\usepackage{ulem}

\templatetype{pnasresearcharticle} 

\title{Narrative structure of A Song of Ice and Fire creates a fictional world with realistic measures of social complexity}

\author[a]{Thomas Gessey-Jones}
\author[b,c]{Colm Connaughton}
\author[d]{Robin Dunbar}
\author[e,f,1]{Ralph Kenna}
\author[g]{P{\'{a}}draig MacCarron}
\author[e]{Cathal O'Conchobhair} 
\author[e,f]{Joseph Yose}

\affil[a]{Fitzwilliam College, University of Cambridge, Storey's Way, Cambridge, CB3 0DG, United Kingdom}
\affil[b]{Mathematics Institute, University of Warwick, Coventry, CV4 7AL, United Kingdom}
\affil[c]{London Mathematical Laboratory, 8 Margravine Gardens, London, W6 8RH, United Kingdom}
\affil[d]{Department of Experimental Psychology, University of Oxford, Anna Watts Building, Radcliffe Observatory Quarter, Oxford OX2 6GG, United Kingdom}
\affil[e]{Centre for Fluid and Complex Systems, Coventry University, Coventry, CV1 5FB, United Kingdom}
\affil[f]{${\mathbb L}^4$ Collaboration \& Doctoral College for the Statistical Physics of Complex Systems, Leipzig-Lorraine-Lviv-Coventry, Europe}
\affil[g]{University of Limerick, Limerick, V94 T9PX, Ireland}

\leadauthor{Gessey-Jones} 

\significancestatement{
We use mathematical and statistical methods 
to probe how a sprawling, dynamic, complex narrative of massive scale achieved broad accessibility and acclaim without surrendering to the need for reductionist simplifications. Subtle narrational tricks such as how natural social networks are mirrored and how significant events are scheduled are unveiled. The narrative network matches evolved cognitive abilities to enable complex messages be conveyed in accessible ways while story time and discourse time are carefully distinguished  in ways matching theories of narratology. 
This marriage of science and humanities opens new avenues to comparative literary studies. It provides quantitative support, for example, for the widespread view that deaths appear to be randomly distributed throughout the narrative even though, in fact, they are not.
}

\authorcontributions{
CC, TGJ, PMC, JY, performed the analysis,
COC gathered the data,
RD, RK designed the research,
everyone contributed to writing of the paper.
}
\authordeclaration{The authors declare that we have no competing interests.}
\correspondingauthor{\textsuperscript{1}To whom correspondence should be addressed. E-mail: Ralph.Kenna@coventry.ac.uk}


\begin{abstract}
Network science and data analytics
are used to quantify static and dynamic  structures in George R.R. Martin's epic novels, {\textit{A Song of Ice and Fire}},  works noted for their scale and complexity. 
By tracking the network of character interactions as the story unfolds, it is found that structural properties remain approximately stable and comparable to real-world social networks. 
Furthermore, the degrees of the most connected characters reflect a cognitive limit on the number of concurrent social connections that humans tend to maintain. 
We also analyse the  distribution of time intervals  between significant deaths measured with respect to the in-story timeline. 
These are consistent with power-law distributions commonly found in inter-event times for a range of non-violent human activities in the real world. 
We propose that structural features in the narrative that are reflected in our actual social world help readers to follow and to relate to the story, despite its sprawling extent. 
It is also found that the distribution of intervals between significant deaths in chapters is different to that for the in-story timeline; 
it is geometric rather than power law. 
Geometric distributions are memoryless in that the time since the last death does not inform as to the time to the next. 
This provides measurable support for the widely held view that significant deaths in {\textit{A Song of Ice and Fire}} are unpredictable chapter-by-chapter. 
\end{abstract}

\dates{This manuscript was compiled on \today}
\doi{\url{www.pnas.org/cgi/doi/10.1073/pnas.XXXXXXXXXX}}

\begin{document}

\maketitle
\thispagestyle{firststyle}
\ifthenelse{\boolean{shortarticle}}{\ifthenelse{\boolean{singlecolumn}}{\abscontentformatted}{\abscontent}}{}


\dropcap{A}{\textit{Song of Ice and Fire}}  is a series of fantasy books written by George R.R. Martin. 
The first five books are
{\textit{A Game of Thrones}}, {\textit{A Clash of Kings}}, {\textit{A Storm of Swords}}, {\textit{A Feast for Crows}} and {\textit{A Dance with Dragons}}. 
Since publication of the first book in 1996, the series has sold over 70~million units 
and has been translated into more than 45~languages.
Martin, a novelist and experienced screenwriter, conceived the sprawling epic as an antithesis to the constraints of film and television budgets. 
Ironically the success of his books attracted interest from  film-makers and television executives worldwide, eventually leading to the television show {\textit{Game of Thrones}}, which first aired in 2011.

Storytelling is an ancient art-form which plays an important mechanism in social bonding \cite{wiessner_embers_2014,dunbar2014mind,KMCMC2017}.
It is recognised that the social worlds created in narratives  often adhere to a ``principle of minimal difference'' whereby social relationships  reflect those in real life --- even if set in a fantastical or improbable world  \cite{palmer_storyworlds_2010}. 
By implication, a social world in a narrative should be constructed in such a way that it can be followed cognitively \cite{dunbar_cognitive_2017}. 
However, the role of the modern story-teller extends beyond the creation of a believable social network. 
As well as an engaging discourse, the manner in which the story is told is important, over and above a simple narration of a sequence of events. 
This distinction is rooted in theories of narratology 
advocated by Schklovsky and Propp 
\cite{new38} and  developed by  Metz, 
Chatman,  Genette  and others  
\cite{Metz,Genette, chatman_story_1980}.

Graph theory has been used to compare character networks to real social networks \cite{newman_why_2003} in mythological \cite{EPL}, Shakespearean \cite{Stiller} and fictional literature \cite{MM}.
To {investigate} the success of \textit{Ice and Fire}, we go beyond graph theory to explore cognitive accessibility as well as differences between how {significant} events are presented  and how they unfold  \cite{RD1}. 
A distinguishing feature of {\textit{Ice and Fire}} is that character deaths are perceived by many readers as random and unpredictable.
Whether you are  ruler of the Seven Kingdoms, heir to a 
ancient dynasty or Warden of the North, your end may be nearer than you think. 
Robert Baratheon met his while boar hunting, Viserys Targaryen while feasting and  Eddard Stark when confessing a crime in an attempt to protect his children.
Indeed, ``Much of the anticipation leading up to the final season (of the TV series) was about who would live or die, and whether the show would return to its signature habit of taking out major characters in shocking fashion'' \cite{VineyardNYT}.
Inspired by this feature, we are  particularly interested in death as signature events in  {\textit{Ice and Fire}} and therefore we study intervals between them \cite{Silvio}. 
To do this, we recognise an important distinction between \textit{{story} time} and \textit{{discourse} time}. 
Story time refers to the order and pace of events as they occurred in the fictional world. 
It is measured in days and months, albeit using the fictional {\textit{Westerosi}} calendar in the case of  {\textit{Ice and Fire}}. 
Discourse time, on the other hand, refers to the order and pacing of events as experienced by the reader; it is measured in chapters and pages.

We find the social network portrayed is indeed similar to those of other social networks and remains, as presented, within our cognitive limit at any given stage. We also find that the order and pacing of deaths differ greatly between discourse time and story time. The discourse is presented in a way that appears more unpredictable than the underlying story; 
had it been told following {\textit{Westerosi}} chronology, the perception of random and unpredictable deaths may be much  less ``shocking'' \cite{beveridge_game_2018,moreno1934shall}.
We suggest that the remarkable juxtaposition of realism (verisimilitude),  cognitive balance and unpredictability  is key to the success of the series.


\matmethods{

To perform this investigation  we draw on two data sets.
The first was extracted manually from  {\textit{Ice and Fire}} by carefully reading the text and noting interactions between characters.
To facilitate comparisons to them, we follow methodologies developed for network analyses of medieval epics 
{\cite{EPL, Iceland,Ossian,CGG}} whereby characters are deemed to have interacted if they directly meet each other or it is explicitly clear from the text they knew one another, even if one or both are dead by that point in the story.
(To our knowledge, no automated method currently exists that has been proven to match this manual approach -- see, e.g., \cite{Marcello}.)
From this data set we construct a network of all the characters in \textit{Ice and Fire} who interact with at least one other. 
Characters are identified as nodes and interactions between them identified as edges (links).
We also gathered temporal data on character deaths for inter-event time analysis.

Fig.~\ref{fig-network} presents, for illustrative purposes, a subset of the  network showing the most {predominant} characters.
The SI contains a similar figure showing only those characters still alive at the end of the fifth book (the survivor network). 
``{Predominance}'' for these illustrations is measured by the number of chapters in which a given character 
interacts with at least one other character and nodes are sized accordingly. 
Each character in Fig.~\ref{fig-network} 
{interacts} in at least 40 chapters. 
The full network is far greater in extent. 
The thickness of the various edges represents the strength of links between nodes as the number of times the corresponding pair of characters interact in the narrative.
The figure is therefore a visual representation of the primary characters and their interactions. 
For example, the enduring importance of characters such as Eddard Stark and Robert Baratheon is clear, despite the fact that both perished early in the story.

\begin{figure}
\centering
\includegraphics[width=1.0\linewidth]{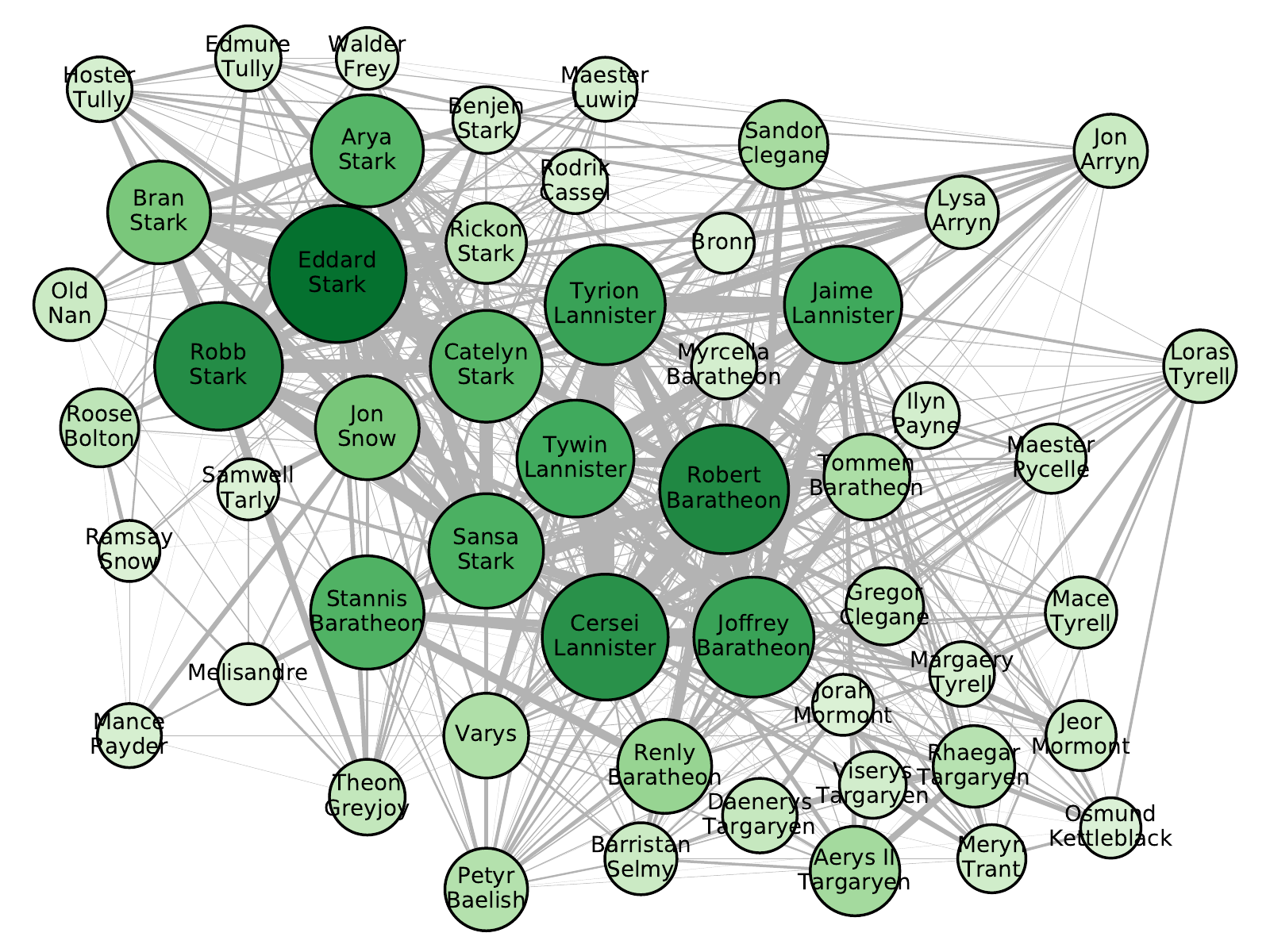}
\caption{Network of the most predominant characters. 
For illustrative purposes we size nodes proportional to the 
number of chapters in which the characters interact. 
Edge thicknesses represent the numbers of times that 
corresponding pair of characters interact in the narrative.}
\label{fig-network} 
\end{figure}

To  analyse  the dynamics and evolution of the narrative we also use a second data set.
This is an approximate timeline of the events of {\textit{Ice and Fire}} indexed by the \textit{Westerosi} calendar date, compiled by fans and followers, and maintained by the \textit{Reddit} user identified as \textit{PrivateMajor}. 
This timeline makes a number of assumptions which are noted in the data set. 
Many of these dates are educated guesses because no explicit in-story timeline is provided by the author. 
According to the \textit{Reddit} timeline, the opening events of {\textit{Game of Thrones}} take place on the 22nd of April of the year 297 and the closing events of {\textit{Dance with Dragons}} take place on the 8th of February in the year 300. 
We used this second set of data to assign an approximate date to each chapter of each book allowing us to study events as they occur within the in-story timeline. 
In many cases, chapters clearly span multiple days. 
In such cases we use the date corresponding to the earliest dated event occurring in that chapter. 
This allows us to order the data in two ways, the order in which the events happen (story time) and the way in which the narrative is told (discourse time).

There are multiple measures of network architecture, nodal importance and edge weights. 
To address the primal issue of societal topology we analyse the full unweighted network. 
We assign a degree to each character as the number of connections it has to other nodes of the network {and we track average values over story and discourse time.} 
Studies have shown that real social networks tend to have properties which distinguish them from other complex networks \cite{newman_why_2003}. 
Notable amongst these is homophily -- the tendency of people to associate with  people who are similar to themselves~\cite{mcpherson_birds_2001}.
One quantitative measure of homophily is  assortativity, the extent to which the degrees of pairs of connected vertices are correlated \cite{newman_assortative_2002}.
A network which has a positive correlation is called assortative and one with a negative correlation is disassortative.

As degree measures how connected a node is, centrality quantifies how close it is to the core of the network.
There are various measures and common examples are betweenness, closeness,  page rank and eigenvector centrality.
We use these tools holistically - no one tool  gives a definitive characterization of verisimilitude or narratology but together they build a picture that we can compare to real networks \cite{newman_why_2003} and to mythological \cite{EPL}, Shakespearean \cite{Stiller} and fictional literature \cite{MM}.
In the next section we present  betweenness, 
which {is a normalized measure of} the number of shortest paths (``geodesics'') between all other nodes that include the particular node in question \cite{freeman_set_1977}.
Nodes with high betweenness are important conduits for information transfer 
and in this sense tend to be more influential.
Further details on data acquisition methodology, network construction and analysis are provided in the SI.
The data and associated analysis codes are available at \cite{github}.

}

\showmatmethods{} 

\section*{Results}

{\textit{Ice and Fire}} is presented from the personal perspectives of 24 ``point of view (POV) characters''.  
A full list of them, ranked by the numbers of chapters from their perspectives, is provided in the SI. 
Of these we consider 14 to be ``major'': 
8 or more chapters,  mostly titled with their names, are relayed from their perspectives.
Tyrion Lannister is major in this sense because the 47 chapters from his perspective 
are titled ``Tyrion I'', ``Tyrion II'', etc.
Arys Oakheart does not meet this criterion as the only chapter related from his perspective is titled ``The Soiled Knight''.
We open this section by reporting how network measures reflect the POV structure.
We then examine the network itself - how it evolves over discourse time, its verisimilitude and the extent to which it  is cognitively accessible. 
Finally we analyse the distributions of time intervals between significant deaths 
and contrast these as measured in story time versus discourse time.


\subsection*{Most important characters}
In networks,  properties such as degree and centrality are signifiers of node importance.
We now rank nodes according to these measures to examine the extent to which they correlate with the POV list.

Table \ref{tab-rankings} lists the 10 characters with the greatest degree and those with the greatest betweenness. 
We present results for the full network (upper panel) and the "survivor" network (lower panel).
The latter contains only those characters possibly still living by the end of the fifth book 
(e.g. a major character whose fate is uncertain by the end of \textit{Dance With Dragons}, is treated as alive). 

The table also lists {major} POV characters that lie outside the top 10 (lower parts of each panel). Degree and betweenness are very different indicators of importance 
than the notion of POV characters. 
However, POV characters form the majority of the top 10  characters when ranked by either measure. There are only three non-POV characters in Table~\ref{tab-rankings}; {Robb Stark}, {Stannis Baratheon} and {Tywin Lannister}. These are highlighted  in bold type. 
The effectiveness of network measures at qualifying character importance is established by the fact that both rankings primarily pick out the POV characters. 
Here we use betweenness as indicative of centrality with other measures
presented in the SI.
Different centrality measures paint similar pictures,  
suggesting the importance-portraits they deliver are quite robust in network  terms.

\begin{table}
\caption{\label{tab-rankings}Characters ranked by various network attributes}
\begin{tabular}{lr}
Full network &   \\
\midrule
Degree & Betweenness Centrality \\
\midrule
1. Jon Snow (214) & 1. Jon Snow (0.0889) \\
2. Jaime Lannister (212) & 2. Barristan Selmy (0.0831) \\
3. Tyrion Lannister (209) & 3. Arya Stark (0.0777) \\
4. Catelyn Stark (204) & 4. Tyrion Lannister (0.0700) \\
5. Arya Stark (192) & 5. Theon Greyjoy (0.0671) \\
6. Theon Greyjoy (175) & 6. Jaime Lannister (0.0606) \\
7. Cersei Lannister (161) & 7. Catelyn Stark (0.0568) \\
\bf{8. Robb Stark (158)} & \bf{8. Stannis Baratheon (0.0519)} \\
9. Sansa Stark (156) & \bf{9. Tywin Lannister (0.0356)} \\
10. Barristan Selmy (156) & 10. Eddard Stark (0.0351) \\
\midrule 
12. Eddard Stark (140) & 12. Sansa Stark (0.0275) \\
16. Brienne of Tarth (108) & 13. Cersei Lannister (0.0250) \\
17. Bran Stark (106) & 14. Brienne of Tarth (0.0236) \\
19. Daenerys Targaryen (104) & 17. Samwell Tarly (0.0207) \\
20. Samwell Tarly (103) & 18. Bran Stark (0.0202) \\
51. Davos Seaworth (72) & 21. Daenerys Targaryen (0.0185) \\
 & 25. Davos Seaworth (0.0167) \\
\bottomrule
\end{tabular}
\begin{tabular}{lr}
Survivor network &   \\
\midrule
Degree & Betweenness Centrality \\
\midrule
1. Tyrion Lannister (162) & 1. Tyrion Lannister (0.0972) \\
2. Jon Snow (150) & 2. Barristan Selmy (0.0952) \\
3. Jaime Lannister (149) & 3. Arya Stark (0.0923) \\
4. Arya Stark (135) & 4. Theon Greyjoy (0.0909) \\
5. Sansa Stark (122) & 5. Jon Snow (0.0871) \\
6. Cersei Lannister (120) & \bf{6. Stannis Baratheon (0.0812)} \\
7. Theon Greyjoy (115) & 7. Jaime Lannister (0.0805) \\
8. Barristan Selmy (103) & 8. Sansa Stark (0.0408) \\
\bf{9. Stannis Baratheon (86)} & 9. Samwell Tarly (0.0320) \\
10. Brienne of Tarth (83) & 10. Cersei Lannister (0.0310) \\
\midrule 
12. Samwell Tarly (79) & 12. Brienne of Tarth (0.0274) \\
18. Daenerys Targaryen (69) & 13. Bran Stark (0.0248) \\
20. Bran Stark (68) & 17. Davos Seaworth (0.0184) \\
38. Davos Seaworth (54) & 33. Daenerys Targaryen (0.0093) \\
\bottomrule
\end{tabular}
\addtabletext{Characters ranked by degree and betweenness centrality (with values in parentheses). 
The three non-POV characters that appear in the top 10 are highlighted in boldface and {major} POV characters who do not appear in the top 10 are also listed. Qualitatively it appears that the 14 {major} POV characters correlate well with the most important characters by both measures.}
\end{table}

\subsection*{Evolution of the social-network structure}

\begin{SCfigure*}[\sidecaptionrelwidth][t]
\centering
\includegraphics[width=11.4cm]{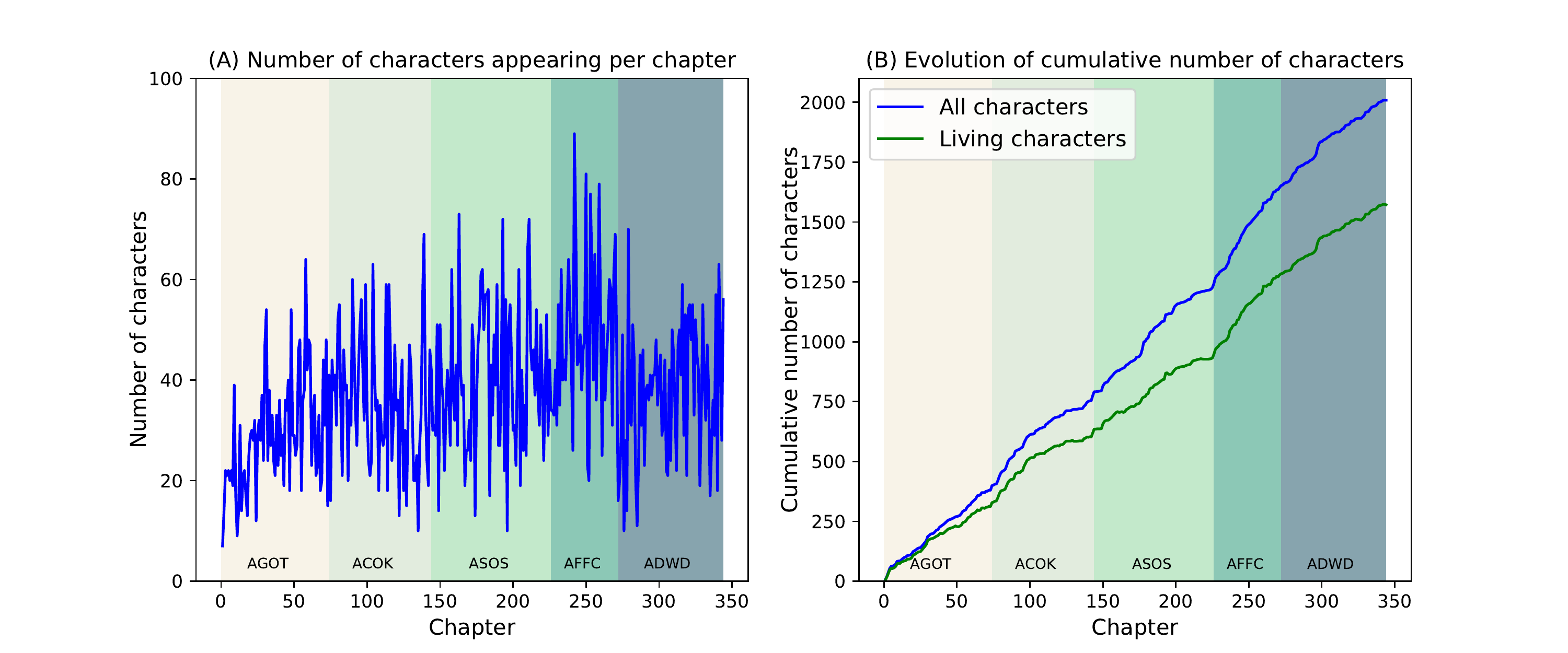} 
\caption{Number of characters in the narrative.
Panel A: Number of characters appearing in each individual chapter.
This shows significant fluctuations chapter by chapter and 
{fluctuates}
around 35 by the end of {{\textit{A Game of Thrones}}.}
Panel B: 
Evolution of the cumulative number of characters appearing in the narrative by chapter (blue)
and of characters introduced who have not yet died (green). 
Both curves grow approximately linearly throughout {\textit{Ice and Fire}}. 
Labels AGOT, ACOK, ASOS, AFFC and ADWD represent 
 {\textit{A Game of Thrones}}, 
     {\textit{A Clash of Kings}},  
    {\textit{A Storm of Swords}}, 
     {\textit{A Feast for Crows}} and
     {\textit{A Dance with Dragons}}, respectively.}
\label{fig-cumulchars_by_chap}
\end{SCfigure*}

From the  five  books containing 343 chapters, 2007 characters were identified, of which 1806 interact with another at least once.
Fig.~\ref{fig-cumulchars_by_chap} depicts how character numbers evolve as the discourse unfolds.
The numbers of characters appearing in each individual chapter are plotted in Panel A. These numbers range from 7 for the first chapter up to 89 for chapter
16 of  {\textit{Feast for Crows}}. 
After a short period of growth in the first book, in which the main characters are introduced, the number of characters per chapter 
settles at around 35. 
This value has been identified as a stable subgrouping within  social networks \cite{hill_social_2003} and as the typical size of (contemporary) bands of hunter-gatherers  
\cite{dunbarnew}.
{It has also been identified as the typical} cast size in Shakespeare's plays~\cite{Stiller} 
 and optimal size for English language and literature research centres  \cite{scientometrics,IMA}.
The cumulative number of characters (introduced up to and including a given chapter) is plotted in blue in Panel~B.
Those who have not explicitly died by the fifth book are depicted in green.
{The near linear growth of each  curve indicates remarkable stability throughout the series.}

\begin{SCfigure*}[\sidecaptionrelwidth][t]
\centering
\includegraphics[width=11.4cm]{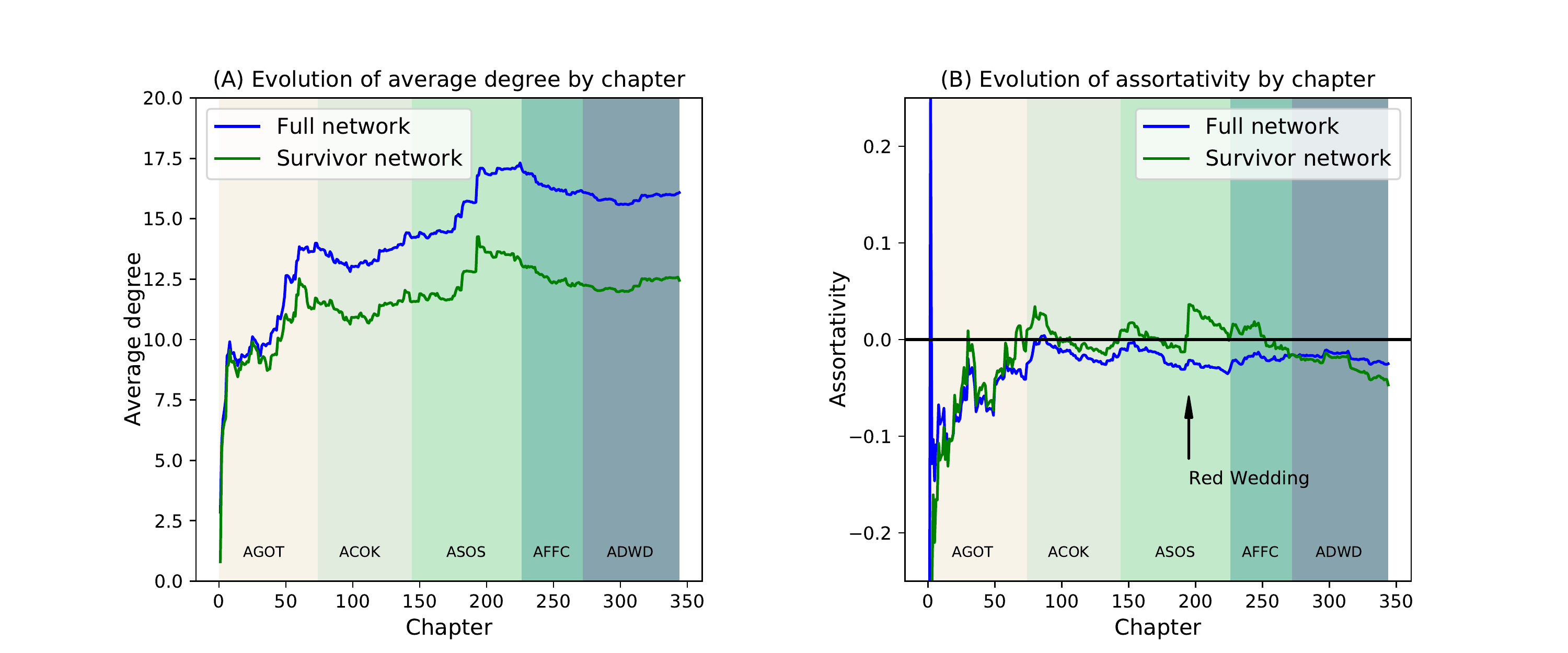} 

\caption{Evolution of network properties by chapter {labelled as in Fig.\ref{fig-cumulchars_by_chap}}.
Panel A: Evolution of the average degree. 
After a period of initial growth as main characters are introduced, the average degree stabilises at around 16 for the network involving  all characters and 
around 12 if only the living characters are included. 
Panel B: Evolution of the degree assortativity. 
After the first book ({\textit{Game of Thrones}}), the assortativity for the living-character networks fluctuate around 0.
While the assortativity of the full network (blue) also fluctuates, it stays slightly disassortative for the later books. }
\label{fig-diagnostics_by_chap}
\end{SCfigure*}

Panel~A of Fig.~\ref{fig-diagnostics_by_chap} 
(which has the same color scheme as  Fig.~\ref{fig-cumulchars_by_chap}) depicts the chapter-by-chapter evolution of the mean degree,  
and Panel~B is the counterpart plot for assortativity. 
The average degree centres around 16 for the full network and around 12 for the survivor network, values that approximate the 15-layer in ego-centric social networks~\cite{zhou_w.-x._discrete_2005,hamilton_complex_2007,mac_carron_calling_2016}.
Although the average degree is small, the distribution is highly skewed as is common in social networks. 
When real social networks are constructed from data, one does not expect every low degree node to necessarily have few connections due to sampling bias.
The same applies to our fictional social network;
since the the narrative is relayed from individual perspectives, the ego networks of POV characters  feature more than those of less prominent characters.
E.g., the highest degree value of 214 belongs to POV character Jon Snow and contrasts markedly with 214 characters having degree 1.
The 14 major POV characters have an average degree of 154.0 within the network of all characters, with standard deviation 47.0. 
This is close to Dunbar's number of 150, the average number of stable relationships usually maintained at any given point in human life 
\cite{dunbarnew}.

Another consequence of the POV style 
is the suppression of assortativity 
compared to  real social networks after the fourth book.
The deflated degrees of the masses relative to POV characters decreases homophily. 
The corollary of this effect is visible in the survivor assortativity jump seen in the third book when Catelyn Stark, an important POV character, is murdered along with some other notable characters at the ``Red Wedding''.
After an initial growth period in the first book,
assortativity fluctuates around 0, before dropping to a slightly negative value (-0.03) by the fifth book.
This is lower than most values measured in real social networks \cite{newman_why_2003} but not by much. 
It is certainly sufficient to endow {\textit{Ice and Fire}} with a greater degree of verisimilitude than more ego-centric networks such as {\textit{Beowulf}} or the {\textit{T{\'{a}}in B{\'{o}} Cuailgne}} \cite{EPL}. 
In comparative-mythological terms, {\textit{Ice and Fire}} has a narrative networks more akin to those of the Icelandic sagas \cite{Iceland}.

Therefore, despite the continuous introduction of new characters, the author has managed to maintain a consistent social network structure.
The number of these interactions is  at the upper end of the cognitive capacity of an average reader. 
Hence while there is a vast number of characters and even greater number of interactions in  {\textit{Ice and Fire}} at any given stage of the narrative, the social network a reader has to consider in order to follow the story is similar in scale to  natural cognitive capacity.

\subsection*{Distributions of inter-event times for significant deaths} 

\begin{SCfigure*}[\sidecaptionrelwidth][t]
\centering
\includegraphics[width=11.4cm]{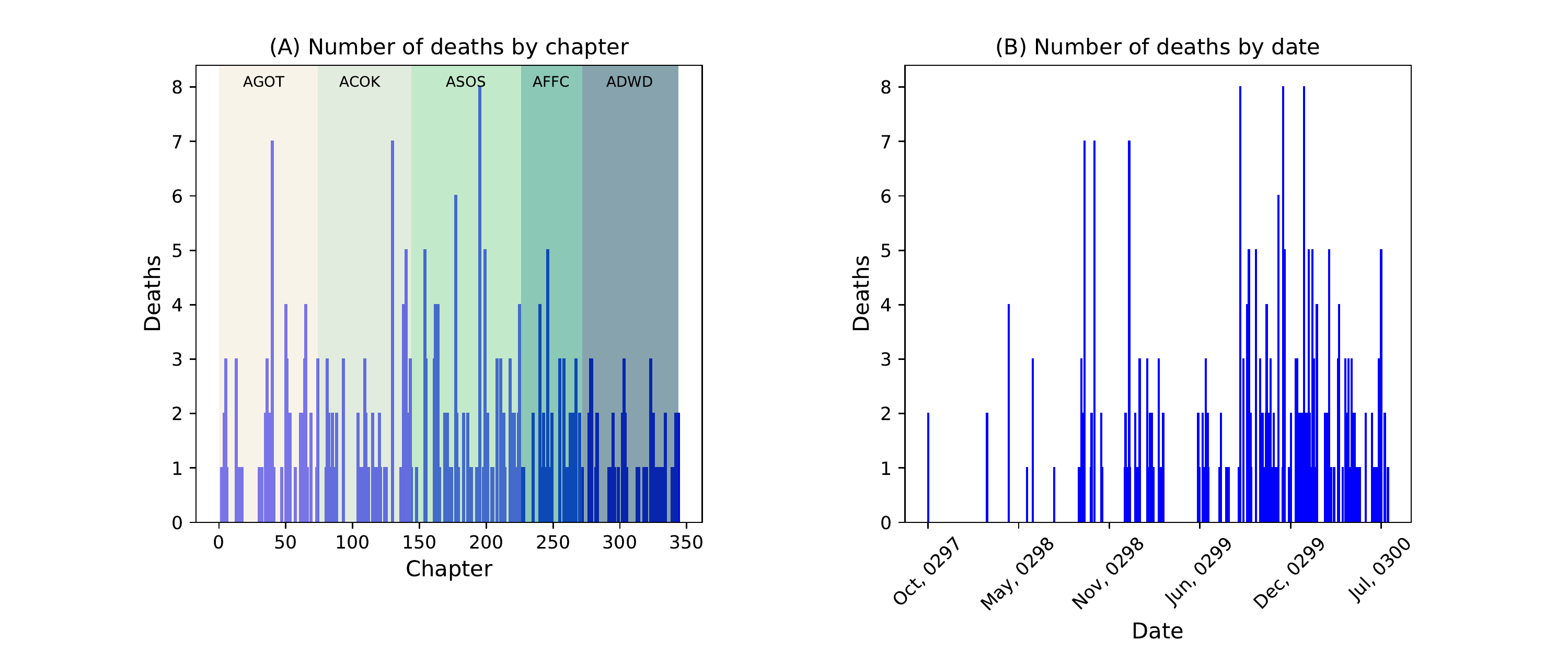}

\caption{Timeline of significant character deaths in  {\textit{Ice and Fire}}.  \label{fig-timeseries_of_deaths}  
Panel~A: Number of deaths by chapter (discourse time).
Panel~B: Number of deaths by date (story time).  }
\end{SCfigure*}

We now turn to consider inter-death story time and inter-death discourse time to reveal an interesting difference between the underlying chronology and how the narrative is presented.
For this purpose we consider only deaths {which we deem to be significant. These are deaths of characters in the network who appear in more then one chapter. We} apply this criterion to avoid the inclusion of the deaths of ``cannon-fodder'' characters whose 
main purpose in the story is to die immediately after they are introduced.
Panel~A of Fig.~\ref{fig-timeseries_of_deaths} shows the number of {significant} character deaths by chapter (discourse time). 
Panel~B gives the same data ordered by date (story time). 
It is striking how deaths appear far more clumped together in story time than in discourse time. The structure of Panel~A helps explain the perception that death can occur unpredictably in the narrative while Panel~B suggests extended ``safe'' periods where no deaths occur. 

These observations can be quantified by examining the empirical distributions of the time intervals between deaths. 
Our data analysis follows  the reasoning described  in \cite{clauset_power-law_2009} and all computations are performed using the associated R package \cite{gillespie_fitting_2015}.  
To explore the (un)predictability of  {\textit{Ice and Fire}} timelines,
we consider the conditional probability that the number of steps (chapters or days) to the next event exceeds 
$n+m$ given that it has already exceeded $m$.
If this is the same as the unconditional probability that the waiting time exceeds $n$,
\begin{equation}
\mathbb{P}(X > n+m\,| \, X > m) = \mathbb{P}(X > n),
\end{equation}
then the inter-event time distribution is said to be memoryless.
In other words, knowing the time since the last event provides  no information about the time to the next event.
It is well known that the geometric distribution 
\begin{equation}
\label{eq-Pgeom}\mathbb{P}(X = n) \equiv P_X(n) = q\,(1-q)^{n-1},
\end{equation}
is the only discrete  inter-event-time distribution that is memoryless and thus maximally unpredictable.
Here $q$ is a parameter to be fitted from data. 
Further details can be found in the SI.


\begin{SCfigure*}[\sidecaptionrelwidth][t]
\centering
\includegraphics[width=11.4cm]{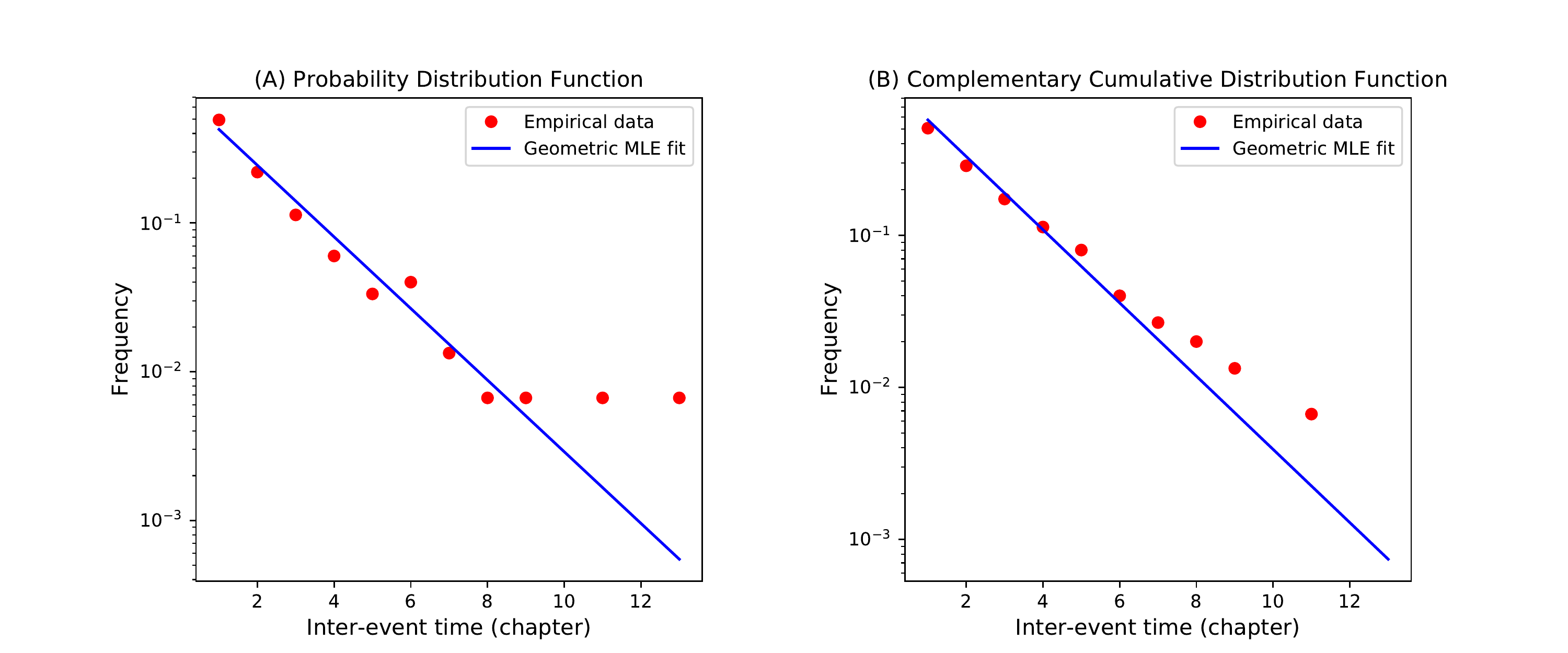}
\caption{Empirical distributions of inter-event times for significant deaths measured by chapter (discourse time), with fit to geometric distribution.  \label{fig-deaths_IET_by_chap}
A geometric distribution is memoryless 
in that it is what would be expected if deaths are maximally unpredictable throughout, as is suggested by many readers/viewers of the series.
}
\end{SCfigure*}

\begin{SCfigure*}[\sidecaptionrelwidth][t]
\centering
\includegraphics[width=11.4cm]{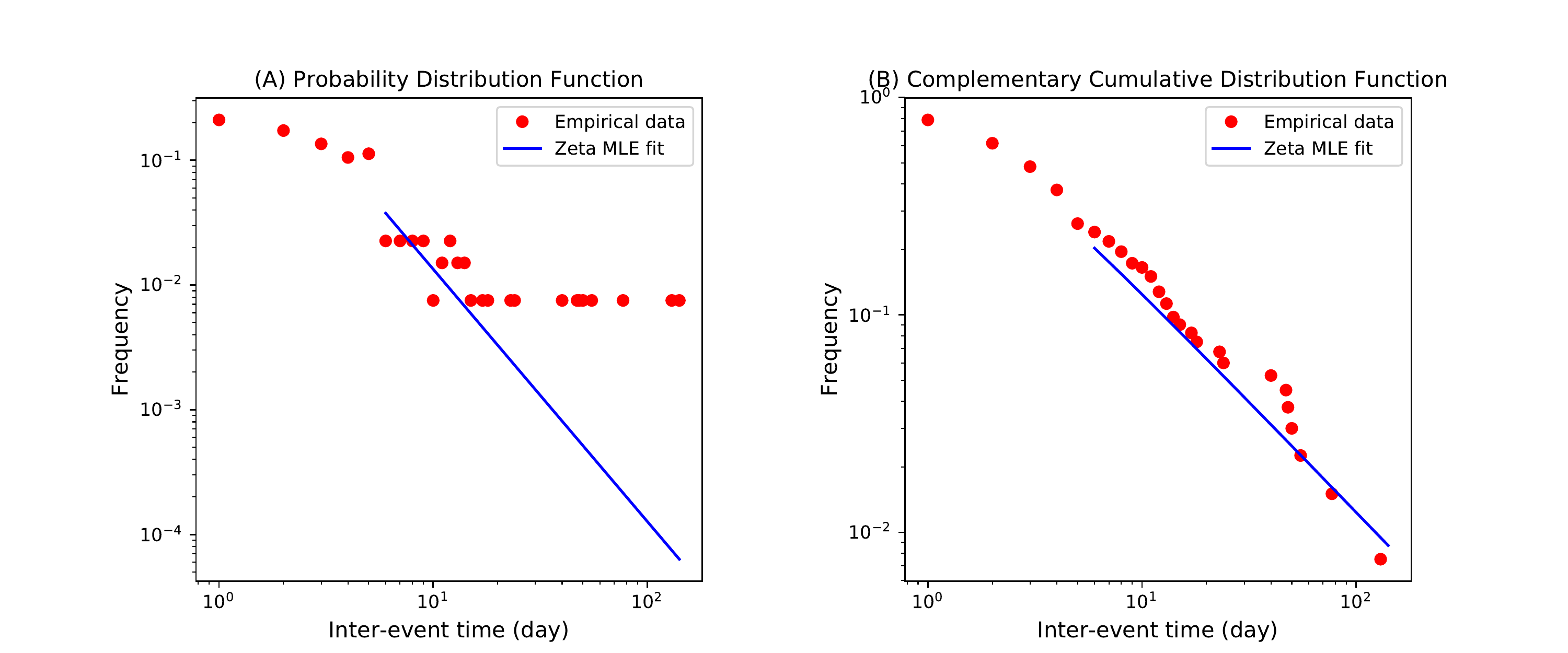}
\caption{ Empirical distributions of inter-event times for significant deaths by date (story time). \label{fig-deaths_IET_by_date} Date here is measured using the fictional Westerosi calendar. Shown in blue is the best-fit discrete power law (Zeta) distribution.}
\end{SCfigure*}

Inter-event discourse time data are presented in Panel A of Fig.\ref{fig-deaths_IET_by_chap} and the corresponding cumulative data are presented in Panel B.
We use the maximum likelihood method to determine the best fits to the geometric distribution for the data and these are also plotted in the figure. 
The associated  p-values characterize goodness of fit, we reject the hypothesis if the p-value is less than 0.05. The results are as follows:\\
\begin{tabular}{ll}
Discourse time: & $q =  0.58$ , $[\ 0.50  ,  0.68\ ]$\\
 & p-value = 0.087;\\
Story time: & $q =  0.12$ , $[\ 0.10  ,  0.15\ ]$\\
 & p-value $\approx 0$.
\end{tabular}\\
Here parenthesized values indicate the approximate 95\% confidence intervals determined by bootstraping.
These results suggest that 
inter-event times for significant deaths in discourse time (chapters) are well described by a geometric distribution and are therefore memoryless.
In contrast, the null hypothesis that significant deaths in story time (calendar) follow a memoryless  geometric distribution can be rejected.
These data are presented in Fig.~(\ref{fig-deaths_IET_by_date}).

Since events in story time are inconsistent with memorylessness, we consider an alternative to the geometric distribution.
Evidence suggests that inter-event time distributions for many (non-violent) human activities in the real world, including communication, entertainment, trading, and work, have power-law tails, usually with exponents between 1 and 2 \cite{barabasi_origin_2005,vajna_modelling_2013,clauset_power-law_2009}. 
Similar heavy tails have been observed in inter-violence intervals 
\cite{simkin2014stochastic,simkin2018statistical} as well as in human behaviour in virtual environments \cite{HMST}. 
Therefore we fit to a discrete power law distribution of the form
\begin{eqnarray}
\label{eq-Pzeta}\mathbb{P}(X = k) \equiv P_X(k) &=& \frac{k^{-\alpha}}{\zeta(\alpha)}.
\end{eqnarray}
Here the exponent $\alpha$ controls the power law and $\zeta(\alpha)$ is the Riemann zeta function. 
The results are as follows:
\begin{tabular}{ll}
Discourse time: & $\alpha = 3.9$, $[2.0, 8.9 \ ]$\\
& $x_0 = 3.7$ , $[\ 1  ,  7\ ]$\\
& p-value = 0.392;\\
Story time: & $\alpha = 2.00$ , $[\ 1.75  ,  2.36\ ]$\\
& $x_0 = 3.6$, $[\ 2 , 8\ ]$\\
 & p-value $\approx 0.428$.
\end{tabular}\\
Again, the uncertainties in parameters indicate approximate 95\% confidence intervals and are estimated by bootstrapping.
The fits, which are plotted for story time in  Fig.~(\ref{fig-deaths_IET_by_date}), 
suggests  that inter-event times for significant deaths by date are indeed well described by a power law distribution with a lower cut-off. 
Interestingly, the story-time exponent $\alpha \approx 2$ is comparable to the values seen in real-world human activities \cite{barabasi_origin_2005,vajna_modelling_2013,clauset_power-law_2009,simkin2014stochastic,simkin2018statistical,HMST}, providing another sense in which the fictitious world of  {\textit{Ice and Fire}} bears quantitative similarity to the real social world. 

At first sight the results for discourse time appear also to suggest that we cannot reject the hypothesis that inter-event times for significant deaths by chapter can  be matched by a power law distribution.
However, the  cutoff $x_0$ excludes long waiting times from the fit.
Unlike the the single-parameter geometric fit, a power-law does not 
match  the entire range of the data. 
(The poorer match of the cut-off power law is also reflected in the uncertainties in the $\alpha$ parameter value which are much higher than their discourse-time counterparts.) 
The memoryless geometric model is therefore the superior description of inter-event times in discourse.

In summary, the inter-event time distribution for {significant deaths by} discourse time is well fitted by a geometric distribution indicating that {such} events  can seem to the reader to occur almost at random intervals. 
However, when analysing deaths in terms of story time, this is not the case, with significant events occurring in a more ``natural'' way. 
Portraying significant events by discourse time instead of as they happen appears to maintain the reader's suspense.

\section*{Discussion}
\label{IV}

{\textit{A Song of Ice and Fire}} is a prodigious modern epic of considerable complexity that remains accessible to a vast congregation of devotees.
Amongst its appeals are the uncertainty and unpredictability of its storyline as characters, including important ones, can be killed off seemingly at random.
{Indeed, not even the POV characters are guaranteed safe passage from one book to the next.}
Here we have shown that the network properties of the society described in  {\textit{Ice and Fire}} are  close to what we expect in real social networks \cite{EPL}.
Also, by relating the story from the points of view of different characters, the total number of interactions at any given stage remains within the average reader's cognitive limit, making it possible to keep track of these relationships.
\cite{RD1}.

The positioning of this paper relative to the context, initiatives and aspirations of digital humanities merits further comment.
The recent review \cite{review} identified some of its methods and themes,
developed in the context of comparative mythology and traditional epic narrative cycles in particular \cite{KMCMC2017}, 
as one of four focal points in the extraction and analysis of character networks
(the remaining three foci being literary analysis, video narratives and computer-science methods aimed at data extraction). 
Here we go beyond such character-network considerations by introducing two new elements to quantitative narratology, namely the questions of cognitive limits and the interplay between story time and discourse time \cite{dunbar_cognitive_2017}. 
Unlike historical, quasi-historical or mythological
chronicles of societies and events, 
a key requirement for fictional storytelling in  {\textit{Ice and Fire}} is that it not spin out of control because of its enormous scale. 
Fictional narratives
require widespread engagement for commercial success. 
Whatever the storyteller's cognitive competences may be \cite{dunbar2014mind}, he has to avoid overtaxing his reader's ability to keep track of the action -- itself related to the number of characters involved  \cite{dunbar_cognitive_2017}. 
If the story is allowed to become too complex, there is a threat of the average reader becoming cognitively overwhelmed  and the story becoming chaotic and unfathomable \cite{dunbar_why_2005}.
 {\textit{Ice and Fire}} avoids this; although more than 2000 characters appear, readers and TV audiences alike remain avidly engaged.

The findings reported here suggest that this is facilitated 
by clever structuring such that each chapter is told 
by different POV characters,  endowed with social networks 
containing only around 150 individuals. 
Moreover, there are only 14 major POV characters. 
These are frequent numbers in the structure of real social networks \cite{hill_social_2003,dunbarnew,zhou_w.-x._discrete_2005,hamilton_complex_2007,dunbar_structure_2015,mac_carron_calling_2016,scientometrics} and they allow the reader to work within natural templates;
the story reflects experiences in the everyday social world, and therefore does not overtax cognitive abilities that are evolved to match these scales \cite{dunbar_cognitive_2017}. 

Our findings on the constraints on the size and structure of the cast of characters are not peculiar to this particular drama. 
Similar numerical constraints have been reported for  Shakespeare's plays \cite{Stiller} and contemporary films  \cite{new33}.
Much of this seems to reflect natural limits on mentalising competences -- the cognitive skills that underpin our ability to handle social relationships in the virtual mental sphere  of the everyday social world \cite{dunbar2014mind,dunbar_cognitive_2017}. These are limited to five orders of intentionality and provide the base from which the scaled layers of social networks are built up \cite{dunbar2014mind}; more importantly, neuroimaging studies have shown that competences in this respect correlate directly with the number of individuals in the 15-layer \cite{lewis_ventromedial_2011,powell_joanne_orbital_2012}. 
That this is important for storytelling has been demonstrated by a series of experimental studies showing that enjoyment of a story is greatest when the number of levels of mentalising (effectively the number of characters involved in a scene) is closest to the reader's own mentalising abilities \cite{carney_inference_2014}. 
Krems et al. \cite{new33} showed that Shakespeare, at least, seemed to adjust the number of characters in a scene {to the effect of remaining within} the mentalising capacities of the audience.

Also, the characteristic unpredictability of the narrative appears in discourse time only, with associated inter-event times {for significant deaths} well described by a memoryless distribution.
In story time the plot unfolds in an altogether different manner for, chronologically, many characters die in a way consistent with regular human activities.
The difference suggests that the author structures the order and pacing of significant events (consciously or subconsciously) to make the series more unpredictable. 
The distinction between story time and discourse time was first identified in the early part of the twentieth century by the influential Russian formalist literary theorists, notably Schklovsky and Propp  \cite{new38}. Their distinction between \textit{fabula} (story, or chronological sequence of events) and \textit{sjuzhet} (plot) is essentially that which we draw here. Schklovsky, in particular, emphasised the importance of "defamiliarisation" (decoupling the temporal sequence of the plot from the chronological storyline) as a device for engaging the reader in the story. Our analysis of significant deaths highlights how effectively Martin exploits this technique.
{Although the question whether the quantitative distinction between story time and discourse time applies to less significant events remains open.}

Thus two important, but conflicting, requirements of effective storytelling are successfully married in  {\textit{Ice and Fire}}: 
(1) the reader's attention is maintained by the unexpected sequencing of {significant} events to encourage page turning to find out why something happened or what happens next;
(2) the reader's sense of what is natural is not overtaxed (i.e. seemingly random events make sense).
This remarkable marriage of verisimilitude (realism) and unpredictability (memorylessness) is achieved in a cognitively engaging manner.

In summary, we show that despite its massive scale, {\textit{Ice and Fire}} is very carefully structured so as not to exceed the natural cognitive capacities of a wide readership. 
Despite its dynamic, extended temporal basis, the structure of its social world mirrors that of natural social networks in ways likely to minimise the cognitive burden on the reader. 
At the same time, the storyteller has manipulated the timeline of the story in such a way as to make it continuously more appealing by making
{significant} events seem random so as to heighten the reader's engagement.
{The identification of patterns of verisimilitude, cognition and unpredictability through computational methods may inspire wider quantitative approaches to other areas of literary study, including drama, television, film, periodicity, genre, canonicity, literature, history, and fantasy.}

\acknow{TGJ was supported by the Gold Travel Fund from Fitzwilliam College and COC by a Coventry University grant. RK, PMC and JY were part supported by the European Commission's 
Marie Curie Action ``International Research Staff Exchange Scheme'' project number 612707 and the Leverhulme Trust research grant ``Women, conflict and peace: gendered networks in early medieval narratives'' RPG-2018-014. PMC was additionally supported by the European Research Council (grant 802421).
RK was additionally supported by a Coventry University City of Culture award.
We thank James Carney 
for discussion.}

\showacknow{} 

\bibliography{pnas-ref}

\end{document}



\maketitle

\SItext

\section*{Data acquisition and network construction}

\subsection{Discourse-time data}
The study reported in the main text is of the first five books of {\textit{A Song of Ice and Fire}}. 
A sixth book, {\textit{The Winds of Winter}}, is currently being written \cite{grrm_wow_exert} and a seventh, titled {\textit{A Dream of Spring}}, is expected to be the concluding volume.
Network data were extracted manually by one of us (O’Conchobhair) carefully reading and  annotating  the books over several months.
Data were entered into a set of spreadsheets, one for each book.
Each spreadsheet was broken down by chapters.
As characters appear within each chapter, new rows were added to the spreadsheet and new columns were added containing the names of characters that they interact with.
When a character appears for the first time in the entire series, this appearance is noted as a ``debut'' in a separate column.
If a character dies in a given chapter, this appearance is noted as a ``death'' in a separate column.
Interactions are deemed meaningful when two characters directly meet or when it is clear from the text that they knew each other, even if one or both is now dead.
Judgement is required to determine when a given interaction is meaningful.
The impossibility of harvesting all meaningful interactions using Natural Language Processing tools is why we prefer a manual data collection protocol over less time-consuming automated approaches.

Additional details on the initial data acquisition process include:
\begin{itemize}
    \item Characters that are not named are not included except in exceptional circumstances when it is clear that namelessness follows from the plot (e.g. the ``White Walkers", the ``Faceless Men") rather than indicating inconsequentiality.
    \item We include characters and interactions that are mentioned in the recollections or stories of others - {what we call} ``historical characters". Most historical characters predate the first book. The debut and death of such characters are recorded as the same chapter in which they are first mentioned.
    \item Multiple instances of a pair of characters interacting within a given chapter are treated as a single entry.
    \item We do not double-count interactions: if character A is recorded as having interacted with character B in a given chapter, we do not record a separate interaction if character B subsequently interacts with character A.
\end{itemize}

 The data sets were then subject to post-processing to correct errors and ensure consistency. 
 In cases of doubts regarding the data four other team members were consulted (Connaughton, Gessey-Jones, MacCarron and Yose) and opinions calibrated through discussion. Consensus assured that no automated calibration method was required \cite{CWC}. Post-processing involved the following checks:
 \begin{itemize}
     \item Standardization of spelling, capitalization and use of white spaces in character names.
     \item Removal of any unintended ``blank'' entries.
     \item Correction of any double-counted interactions.
     \item Accounting for aliasing. 
     For example,  Arstan Whitebeard appears in  \textit{A Clash of Kings} and, over time, the reader learns that this character is none other than Barristan Selmy, a noble Westerosi knight and former leader of the esteemed Kingsguard. 
     Because they are one and the same, Arstan Whitebeard and Barristan Selmy are treated as a single individual.
 \end{itemize}

\subsection{Story-time data}
The data sets described above are indexed by chapter and are readily used to measure the evolution of the narrative in discourse time. 
To  generate the timeline of events in story time, a second data set is used. 
This was compiled by fans and followers of the series and is maintained by the Reddit user identified as PrivateMajor \cite{Reddit}.
It assigns dates in the Westerosi calendar to the main events in the narrative.
Some of these dates are necessarily educated guesses because no explicit in-story timeline is provided by the author. 
We link the two data sets at chapter level. 
I.e., we assign all the events described in each chapter to a single date that is intended to be representative of that chapter.
This means that the granularity of the temporal data extends to chapter level.


\subsection{Network construction}
The analysis contained in the main text is restricted to { characters who interact with at least one other character}. This criteria select 1806 characters from the 2007 unique characters contained in the original data.
{The criterion was chosen to minimise the fragmentation of the network. This is important when we calculate various centrality measures since quantities like betweenness, closeness and eigenvector centrality are only meaningfully applicable to the largest connected component of a network.
It also goes some way to excluding historic characters, who are not involved in the story other then being briefly mentioned.}

\subsection{Reproducibility}
The data and the codes used to generate the analyses, tables and figures contained in this paper are publicly available at \cite{github}.

\section*{Additional network visualisations}

\begin{figure}[t]
\centering
\includegraphics[width=0.5\linewidth]{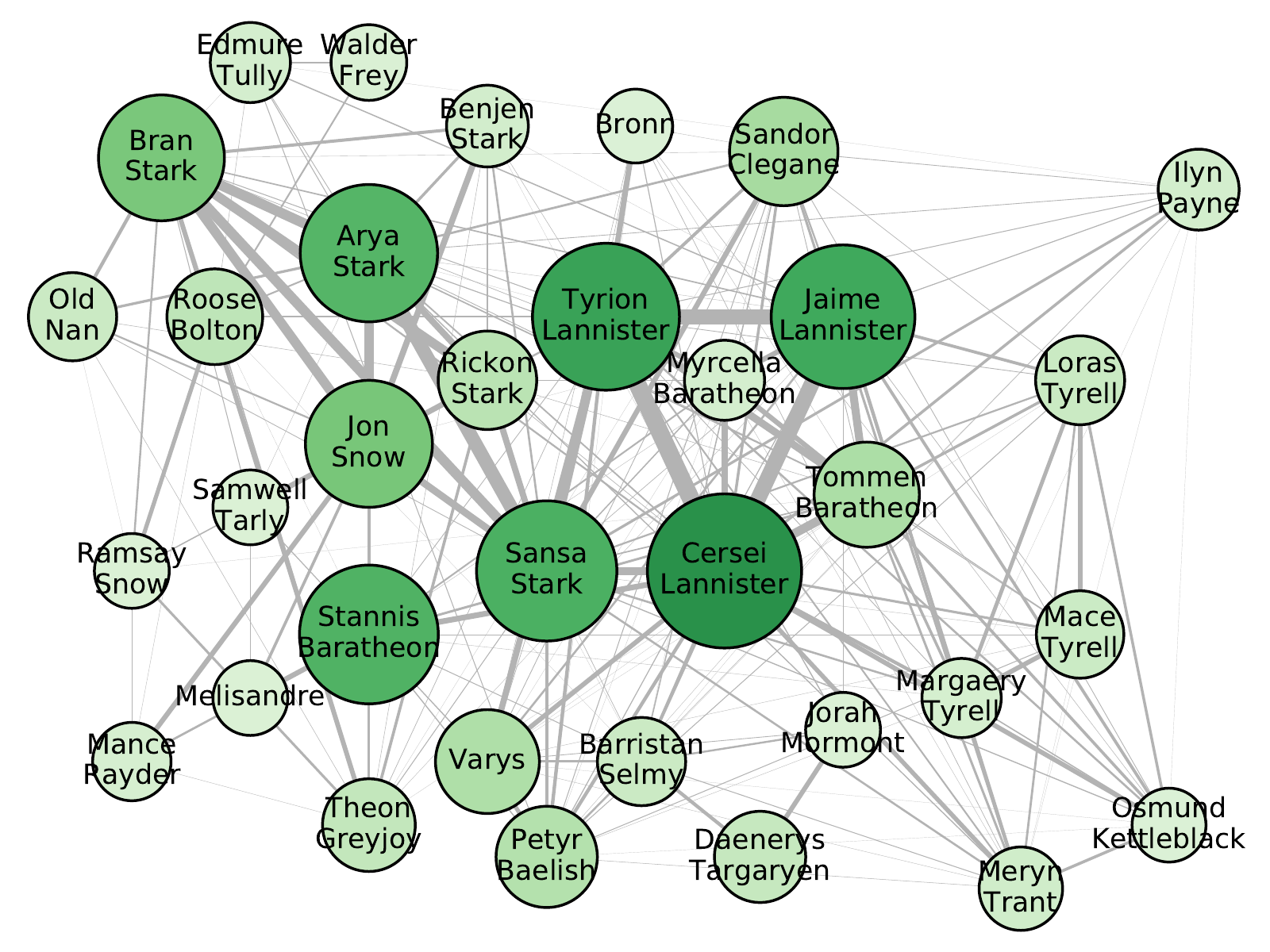}
\caption{Visualisation of the core of the survivor network, i.e., the characters still alive at the end of the fifth book. This sub-network contains the characters who interact with an other in at least 40 chapters, the complete network is much larger.
Node size is proportional to the number of chapters in which a given characters interacts with another. 
Edge thickness is proportional to the number of times that a given pair of characters interact in the narrative.}
\label{fig-network} 
\end{figure}

\begin{figure}[t]
\centering
\includegraphics[width=0.5\linewidth]{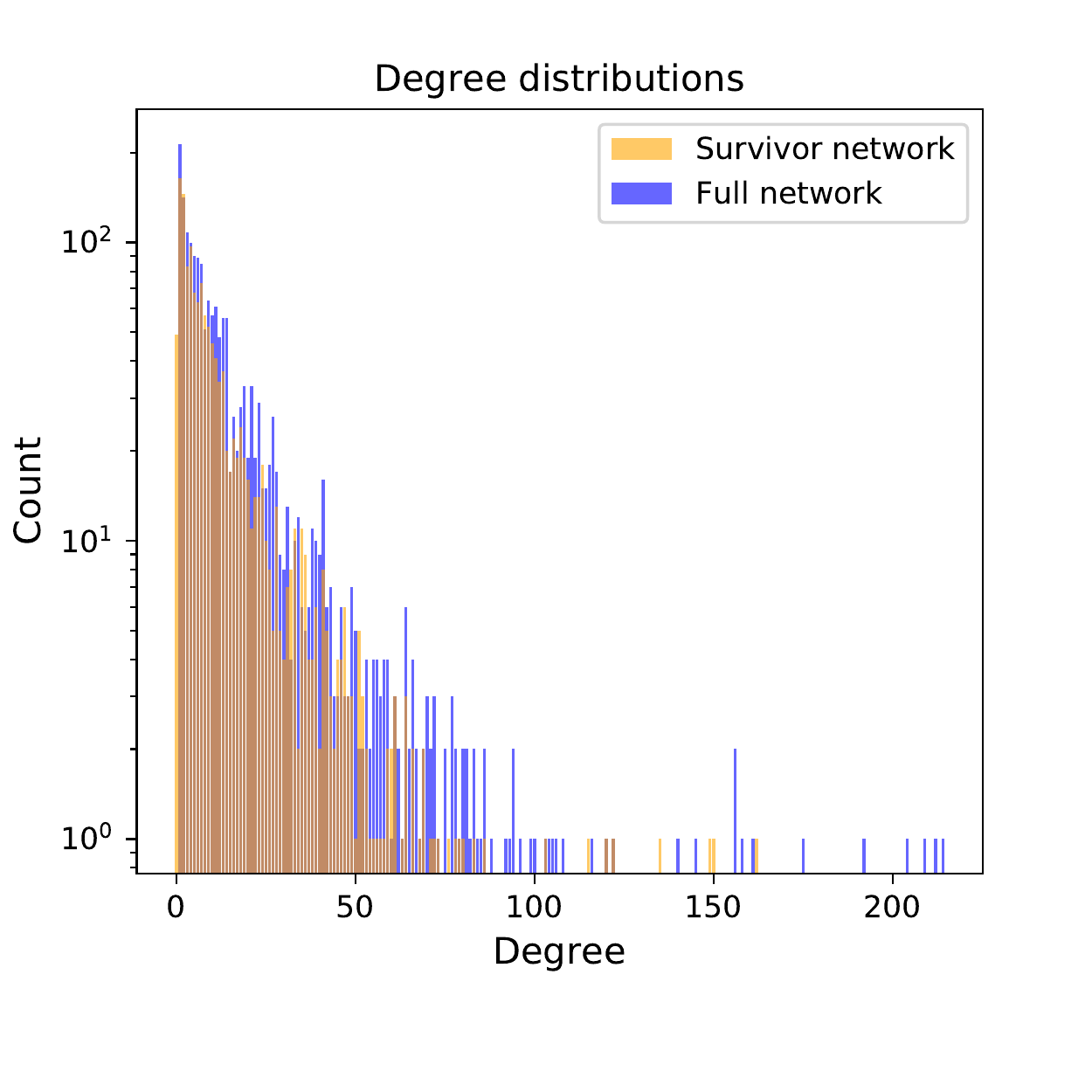}
\caption{Degree distribution for the full network of characters and the survivor network. Here Count is the number of characters with a given degree value. In the full network {214 characters} have degree value one and POV character Jon Snow has the highest degree value of 214.}
\label{fig-degree_distr} 
\end{figure}

We include here some additional network visualisations for completeness. 
 
Figure~\ref{fig-network} shows the most predominant nodes in the survivor network - characters who have not explicitly died by the end of the fifth book. It is the survivor counterpart of Figure 1 of the main text, which includes predominant characters from the full network. As in the main text we determine predominance by the number of chapters in which the character interacts with an other. 

Both networks are qualitatively similar.
Together they illustrate the enduring importance of {characters such as} Eddard Stark and Robert Baratheon to the narrative despite the fact that both characters died in the first book.
 
The degree distributions of the full network and the survivor network are compared in Fig. ~\ref{fig-degree_distr}.
The two are qualitatively similar with a small numbers of very high degree nodes in the tails of both distributions.

\section*{Memoryless distributions}
\label{sec-SI}
We provide here a summary of the mathematics underlying the notion of {\textit{memorylessness}} of inter-event time distributions. 
A distribution of inter-event times is memoryless if knowing the time since the last observed event neither increases nor decreases the probability of observing an event in a future time interval.
These results are well known to statisticians and probability theorists (see, for example, \cite{daley2003introduction}) and are reproduced here solely to render the manuscript self-contained. 
We first describe the continuous case where times between events are real numbers.
In this case, memorylessness is modelled by the exponential distribution.
We then consider the discrete case where times between events are measured in integer units. 
In this case, the memorylessness comes from the geometric distribution, the discrete analogue of an exponential distribution.

Let $X$ be a random variable representing the time between discrete events. In our case, events are the significant deaths of characters in {\textit{Ice and Fire}}. 
If $X$ is exponentially distributed, its probability distribution function is
\begin{equation}
\label{eq-memoryless}
\mathbb{P}(X = x) \equiv P_X(x) = \lambda\,{\rm e}^{-\lambda\,x},
\end{equation}
where $\lambda$ is the rate parameter.
It is the average rate at which events occur, since the mean waiting time is $\mathbb{E}[X] = {1}/{\lambda}$.  
Although there is a well-defined mean waiting time, the distribution function is {\textit{memoryless}} in the sense that knowledge of the time since the last event provides no information about the waiting time to the next event.
To see this, we introduce the cumulative distribution function 
\begin{equation}
\mathbb{P}(X  \leq x) \equiv F_X(x) = 1 - {\rm e}^{-\lambda\,x},
\end{equation}
and its complement, the complementary cumulative distribution function (or survival function),
\begin{equation}
\label{eq-expSurvivalFn}
\mathbb{P}(X  > x) \equiv S_X(x) =  {\rm e}^{-\lambda\,x}.
\end{equation}
Using the law of conditional probability and Eq.~(\ref{eq-expSurvivalFn}), the left-hand side of Eq.~(\ref{eq-memoryless}) becomes
\begin{eqnarray*}
\mathbb{P}(X > t+s\,| \, X > t) &=& \frac{\mathbb{P}(X > t+s)}{\mathbb{P}(X > t)}\\
&=&\frac{S_X(t+s)}{S_X(t)}\\
&=&\frac{{\rm e}^{-\lambda\,(t+s)}}{{\rm e}^{-\lambda\,t}}\\
&=& {\rm e}^ {-\lambda\,s}\\
&=& S_X(s)\\
&=&  \mathbb{P}(X > s).
\end{eqnarray*}
This means that the probability that the waiting time exceeds $t+s$, given that it has already exceeded time $t$, is the same as the unconditional probability that the waiting time exceeds $s$. 
In other words, knowing that the event has not occurred for a time $t$ neither increases nor decreases the probability that one will wait a further time $s$ before the next event. 
Exponentially distributed waiting times are thus maximally unpredictable in the sense that knowing the time since the last event provides  no information about the time to the next event.

However, the exponential distribution is continuous and waiting times between significant events in the narrative take integer values --- chapters or days. 
We therefore consider the discrete analogue of the exponential distribution, namely the geometric distribution. 
This is the distribution of the number, $X$, of Bernoulli trials needed to deliver one event where the probability of success in a single trial is $q$. 
This distribution is supported on the positive integers. 
The corresponding probability {distribution} function, cumulative probability distribution function  and complementary cumulative {distribution} function  are given respectively by:
\begin{eqnarray}
\label{eq-Pgeom}\mathbb{P}(X = n) \equiv P_X(n) &=& q\,(1-q)^{n-1},\\
\mathbb{P}(X \leq n) \equiv F_X(n) &=& 1 - (1-q)^{n},\\
\mathbb{P}(X > n) \equiv S_X(n) &=&  (1-q)^{n}.
\end{eqnarray}
A similar calculation to the one above illustrates that the geometric distribution also has the memoryless property, Eq.~(\ref{eq-memoryless}). 
Indeed, the exponential and geometric distributions are the only distributions with this property and they share a single functional form through the relation $\lambda = - \log (1-q)$.

\section*{Point of view (POV) characters}

In Table \ref{tab-POV} we present the POV characters ranked according to the number of chapters narrated from their perspectives. 
This list {\textit{excludes}} prologues and epilogues which are usually narrated from the perspective of an insignificant character who often dies at the end of their chapter. 
We draw a distinction between major and minor POV characters. 
The former are named in {most of} their chapter title and have at least 8 chapters attributed to them. 
The latter are rarely named in their chapter titles, instead often referred to by descriptors, and are associated with fewer chapters - 4 at most.
Independently of chapter titles, the gap between 4 and 8 chapters also makes this distinction plausible on numerical grounds.

\begin{table}
\centering
\caption{\label{tab-POV}POV characters ranked by number of chapters}
\begin{tabular}{ll|ll}
Major POV characters & &Minor POV characters &  \\
Name & Chapters & Name & Chapters  \\
\midrule
Tyrion Lannister& 47& Asha Greyjoy& 4\\
Jon Snow& 42& Victarion Greyjoy& 4\\
Arya Stark& 33& Quentyn Martell& 4\\
Daenerys Targaryen& 31& Barristan Selmy& 4\\
Catelyn Stark& 25& Aeron Greyjoy& 2\\
Sansa Stark& 24& Areo Hotah& 2\\
Bran Stark& 21& Arianne Martell& 2\\
Jaime Lannister& 17& Jon Connington& 2\\
Eddard Stark& 15& Arys Oakheart& 1\\
Davos Seaworth& 13& Melisandre& 1\\
Theon Greyjoy& 13& & \\
Cersei Lannister& 12& & \\
Samwell Tarly& 10& & \\
Brienne of Tarth& 8& & \\
\bottomrule
\end{tabular}
\addtabletext{POV characters ranked by number of chapters.}
\end{table}

\section*{Different network measures of character centrality}
\label{sec-SI_network_properties}
In Table \ref{tab-rankings} we compare the most highly ranked characters in {\textit{Ice and Fire}} according to several  centrality measures:
\begin{itemize}
    \item Betweenness centrality \cite{newman2018networks}: the {normalized} number of shortest paths between pairs of nodes that pass through a given node.
    \item Closeness centrality \cite{bavelas1950communication,newman2018networks}: the reciprocal of the average distance of a node from all the other nodes on the graph.
        \item Eigenvector centrality \cite{newman2008mathematics,langville2005survey}: a measure of relative importance where the importance of a node is the average of the importance of its immediate neighbours. A node with high eigenvector centrality is connected to many nodes which themeselves have high eigenvector centrality.
    \item Page rank \cite{page1999pagerank,langville2005survey}: a ranking based on the proportion of time a random walker on the graph spends visiting any given node. Mathematically this is a variation of eigenvector centrality in which the total contribution of any node to the scores of its neighbours is 1, regardless of how many neighbours that node has.
\end{itemize}

In Table \ref{tab-rankings} we list the top 15 characters by each of these measures of importance for both the full network (upper sub-table) and the survivor network (lower sub-table). 
Non-POV characters that appear in the top 15 are indicated in boldface.
The ranks of any major POV characters that do not appear in the top 15 are also shown below the line in each case. We can draw several conclusions from these rankings
\begin{itemize}
    \item With the exception of eigenvector centrality, each of these choices of centrality picks out comparable rankings of important characters, with the POV characters heavily represented. This indicates that our analysis is robust.
    \item The character Daenerys Targaryen stands out by how low her ranking appears in terms of betweenness and centrality.
    This is the case despite her being a central character to the story and major POV character. It reflects an important aspect of the plot: for all of the first five books, Daenerys spends her time on a different continent, far from most of the rest of the action. She is distanced from the other main characters both in network terms and in terms of plot.
    \item The distinctiveness of the rankings by eigenvector centrality is striking and reflects an interesting intersection of a feature of the plot with a feature of the algorithm. Note how the list of highest eigenvector centrality characters in the survivor network is dominated by members of House Frey. Walder Frey, the head of House Frey, is an important villain in the narrative, noted for his fecundity and his devastating betrayal of House Stark at the Red Wedding. His many offspring form a large, highly interconnected clique within the character network. This  structure results in high eigenvector centrality; the best way to increase eigenvector centrality is to form many reciprocal links within a small group since this both increases the number of links to a node and increases the relative number of links of these neighbours. Interestingly, this ability to ``game" the eigenvector score by cross-linking is one of the key problems that was addressed by Brin and Page in designing the Page Rank algorithm for ranking web pages that led to the creation of the Google search engine.  
    \item We note that the page rank measure correlates most closely with the POV characters in both the full and survivor networks. Page Rank is not skewed by the strong cross-linking of Frey clique because the total contribution of a node to the score of its neighbours is always one: a ``one node - one vote'' model. Excessive cross-linking within a clique therefore only serves to dilute the score. Fans of the narrative may find it  interesting to note that the Page Rank score is also the only one to identify the importance of Daenerys Targaryen.
\end{itemize}

\begin{table}
\centering
\caption{\label{tab-rankings}Characters ranked by different centrality measures}
\begin{tabular}{lccr}
All characters & & &  \\
\midrule
Betweenness Centrality & Closeness Centrality & Page Rank & Eigenvector Centrality  \\
\midrule
1. Jon Snow (0.0889) & 1. Jaime Lannister (0.4007) & 1. Arya Stark (0.0072) & 1. Catelyn Stark (0.1814) \\
2. Barristan Selmy (0.0831) & 2. Tyrion Lannister (0.3943) & 2. Jon Snow (0.0067) & 2. Jaime Lannister (0.1583) \\
3. Arya Stark (0.0777) & 3. Catelyn Stark (0.3909) & 3. Tyrion Lannister (0.0061) & \bf{3. Robb Stark (0.1547)} \\
4. Tyrion Lannister (0.0700) & 4. Eddard Stark (0.3900) & 4. Jaime Lannister (0.0060) & 4. Tyrion Lannister (0.1451) \\
5. Theon Greyjoy (0.0671) & \bf{5. Robert Baratheon (0.3895)} & 5. Theon Greyjoy (0.0056) & 5. Cersei Lannister (0.1392) \\
6. Jaime Lannister (0.0606) & 6. Arya Stark (0.3855) & 6. Barristan Selmy (0.0052) & 6. Sansa Stark (0.1350) \\
7. Catelyn Stark (0.0568) & \bf{7. Stannis Baratheon (0.3854)} & 7. Catelyn Stark (0.0051) & \bf{7. Tywin Lannister (0.1247)} \\
\bf{8. Stannis Baratheon (0.0519)} & \bf{8. Joffrey Baratheon (0.3815)} & 8. Cersei Lannister (0.0043) & \bf{8. Joffrey Baratheon (0.1198)} \\
\bf{9. Tywin Lannister (0.0356)} & 9. Sansa Stark (0.3804) & \bf{9. Tywin Lannister (0.0042)} & 9. Eddard Stark (0.1139) \\
10. Eddard Stark (0.0351) & 10. Jon Snow (0.3755) & 10. Eddard Stark (0.0040) & \bf{10. Hosteen Frey (0.1096)} \\
\bf{11. Robert Baratheon (0.0288)} & 11. Cersei Lannister (0.3725) & 11. Sansa Stark (0.0040) & \bf{11. Walder Frey (0.1089)} \\
12. Sansa Stark (0.0275) & \bf{12. Tywin Lannister (0.3720)} & \bf{12. Stannis Baratheon (0.0038)} & \bf{12. Petyr Baelish (0.1079)} \\
13. Cersei Lannister (0.0250) & \bf{13. Robb Stark (0.3719)} & \bf{13. Robb Stark (0.0038)} & \bf{13. Walder Rivers (0.1068)} \\
14. Brienne of Tarth (0.0236) & 14. Barristan Selmy (0.3702) & 14. Daenerys Targaryen (0.0036) & \bf{14. Edmure Tully (0.1055)} \\
\bf{15. Jorah Mormont (0.0227)} & 15. Theon Greyjoy (0.3694) & 15. Brienne of Tarth (0.0034) & \bf{15. Tommen Baratheon (0.1027)} \\
\midrule 
17. Samwell Tarly (0.0207) & 17. Bran Stark (0.3640) & 18. Samwell Tarly (0.0031) & 19. Arya Stark (0.0982) \\
18. Bran Stark (0.0202) & 48. Brienne of Tarth (0.3319) & 21. Bran Stark (0.0030) & 32. Theon Greyjoy (0.0880) \\
21. Daenerys Targaryen (0.0185) & 76. Samwell Tarly (0.3234) & 24. Davos Seaworth (0.0029) & 55. Bran Stark (0.0774) \\
25. Davos Seaworth (0.0167) & 166. Davos Seaworth (0.3033) &  & 67. Jon Snow (0.0693) \\
 & 227. Daenerys Targaryen (0.2973) &  & 78. Brienne of Tarth (0.0661) \\
 &  &  & 209. Samwell Tarly (0.0261) \\
 &  &  & 326. Davos Seaworth (0.0167) \\
 &  &  & 396. Daenerys Targaryen (0.0125) \\
\bottomrule
\end{tabular}
\begin{tabular}{lcccr}
Surviving characters only &  & & & \\
\midrule
Betweenness Centrality & Closeness Centrality & Page Rank & Eigenvector Centrality \\
\midrule
1. Tyrion Lannister (0.0972) & 1. Jaime Lannister (0.3754) & 1. Arya Stark (0.0090) & \bf{1. Hosteen Frey (0.1693)} \\
2. Barristan Selmy (0.0952) & 2. Tyrion Lannister (0.3633) & 2. Tyrion Lannister (0.0081) & \bf{2. Walder Frey (0.1689)} \\
3. Arya Stark (0.0923) & 3. Arya Stark (0.3617) & 3. Jon Snow (0.0078) & \bf{3. Walder Rivers (0.1684)} \\
4. Theon Greyjoy (0.0909) & \bf{4. Stannis Baratheon (0.3567)} & 4. Jaime Lannister (0.0073) & \bf{4. Edwyn Frey (0.1632)} \\
5. Jon Snow (0.0871) & 5. Sansa Stark (0.3545) & 5. Theon Greyjoy (0.0065) & \bf{5. Black Walder (0.1595)} \\
\bf{6. Stannis Baratheon (0.0812)} & 6. Theon Greyjoy (0.3453) & 6. Barristan Selmy (0.0058) & \bf{6. Lothar Frey (0.1578)} \\
7. Jaime Lannister (0.0805) & 7. Jon Snow (0.3444) & 7. Cersei Lannister (0.0054) & \bf{7. Aenys Frey (0.1564)} \\
8. Sansa Stark (0.0408) & 8. Cersei Lannister (0.3429) & 8. Sansa Stark (0.0053) & \bf{8. Roslin Frey (0.1560)} \\
9. Samwell Tarly (0.0320) & 9. Bran Stark (0.3407) & \bf{9. Stannis Baratheon (0.0046)} & \bf{9. Fat Walda (0.1496)} \\
10. Cersei Lannister (0.0310) & \bf{10. Tommen Baratheon (0.3394)} & 10. Asha Greyjoy (0.0044) & \bf{10. Ami Frey (0.1470)} \\
11. Victarion Greyjoy (0.0292) & 11. Barristan Selmy (0.3295) & 11. Brienne of Tarth (0.0043) & \bf{11. Joyeuse Erenford (0.1465)} \\
12. Brienne of Tarth (0.0274) & \bf{12. Jeyne Poole (0.3277)} & 12. Samwell Tarly (0.0039) & \bf{12. Perwyn Frey (0.1450)} \\
13. Bran Stark (0.0248) & \bf{13. Roose Bolton (0.3271)} & 13. Victarion Greyjoy (0.0039) & \bf{13. Olyvar Frey (0.1435)} \\
\bf{14. Roose Bolton (0.0212)} & \bf{14. Sandor Clegane (0.3263)} & 14. Daenerys Targaryen (0.0036) & \bf{14. Benfrey Frey (0.1393)} \\
15. Asha Greyjoy (0.0197) & \bf{15. Myrcella Baratheon (0.3219)} & 15. Davos Seaworth (0.0036) & \bf{15. Alyx Frey (0.1390)} \\
\midrule 
17. Davos Seaworth (0.0184) & 21. Brienne of Tarth (0.3140) & 18. Bran Stark (0.0034) & 21. Jaime Lannister (0.1246) \\
33. Daenerys Targaryen (0.0093) & 39. Samwell Tarly (0.3017) &  & 40. Tyrion Lannister (0.0947) \\
 & 117. Davos Seaworth (0.2759) &  & 42. Cersei Lannister (0.0916) \\
 & 507. Daenerys Targaryen (0.2511) &  & 45. Sansa Stark (0.0866) \\
 &  &  & 51. Arya Stark (0.0700) \\
 &  &  & 54. Theon Greyjoy (0.0656) \\
 &  &  & 58. Brienne of Tarth (0.0604) \\
 &  &  & 80. Bran Stark (0.0474) \\
 &  &  & 93. Jon Snow (0.0376) \\
 &  &  & 193. Samwell Tarly (0.0185) \\
 &  &  & 247. Davos Seaworth (0.0121) \\
 &  &  & 424. Daenerys Targaryen (0.0053) \\
\bottomrule
\end{tabular}
\addtabletext{Characters ranked by various importance measures (with values in parentheses). The non-POV characters that appear in the top 15 are highlighted in boldface and {major} POV characters who do not appear in the top 15 are also listed. Qualitatively it appears that the 14 \red{major} POV characters correlate well with the most important characters by all measures.}
\end{table}

\bibliography{pnas-ref}